\documentclass[preprint,aps,showpacs,nofootinbib]{revtex4}

\usepackage{epsfig,amssymb,amsmath}

\def\bea{\begin{eqnarray}}
\def\eea{\end{eqnarray}}
\def\bec{\begin{center}}
\def\ec{\end{center}}

\def\beq{\begin{equation}}
\def\eeq{\end{equation}}

\begin{document}
\draft
\tighten
\preprint{KAIST-TH 2008/03}
\preprint{KIAS-P08032}
\preprint{TU-817}
\title{\large \bf Flavor and CP conserving moduli mediated SUSY breaking
               \\ in flux compactification}
\author{
Kiwoon Choi\footnote{email: kchoi@muon.kaist.ac.kr}$^1$, Kwang Sik
Jeong\footnote{email: ksjeong@kias.re.kr}$^2$, Ken-Ichi
Okumura\footnote{email: okumura@tuhep.phys.tohoku.ac.jp}$^3$}
\address{
$^1$Department of Physics, Korea Advanced Institute of Science and
Technology,
Daejeon 305-701, Korea \\
$^2$School of Physics, Korea Institute for Advanced Study,
Seoul 130-722, Korea \\
$^3$Department of Physics, Tohoku University, Sendai 980-8578, Japan}
%\date{\today}
%\maketitle

\vspace{1cm}

\begin{abstract}

In certain class of flux compactification, moduli mediated
supersymmetry (SUSY) breaking preserves flavor and CP at leading
order in the perturbative expansion controlled by the vacuum
expectation value of the messenger modulus.  Nevertheless there
still might be dangerous flavor or CP violation induced by higher
order K\"ahler potential. We examine the constraints on such SUSY
breaking scheme imposed by low energy flavor and/or CP violating
observables. It is found that all phenomenological constraints can
be satisfied even for generic form of higher order K\"ahler
potential and sparticle spectra in the sub-TeV range, under
plausible assumptions on the size of higher order correction and
flavor mixing angles. This implies for instance that mirage
mediation scheme of SUSY breaking, which involves such modulus
mediation together with an anomaly mediation of comparable size, and
also the modulus-dominated mediation realized in flux
compactification can be free from the SUSY flavor and CP problems,
while giving gaugino and sfermion masses in the sub-TeV range.

\end{abstract}

\pacs{} \maketitle

\section{introduction}

Weak scale supersymmetry (SUSY)  is one of the prime candidates for
physics beyond the standard model at the TeV scale
\cite{Nilles:1983ge} which will be probed soon at the LHC. One key
question on weak scale SUSY is the origin of the soft SUSY breaking
terms of  visible gauge and matter superfields in low energy
effective lagrangian \cite{soft}. Those soft terms are required to
preserve flavor and CP with high accuracy, which severely constrains
the possible mediation mechanism of SUSY breaking. There are certain
schemes such as gauge mediation \cite{gauge.mediation} and anomaly
mediation \cite{anomaly.mediation} in which the standard model gauge
interactions play a dominant role for the mediation, thereby
automatically yield flavor and CP conserving soft terms.
On the other hand, it is commonly thought that gravity mediation
\cite{gravity.mediation} generically leads to dangerous flavor
and/or CP violation, therefore needs an additional ingredient  in
order to be consistent with low energy observations.

The messenger scale of gravity mediation is near the Planck scale
$M_{Pl}=2.4\times 10^{18}$ GeV which might be identified as the
scale of quantum gravity. As string theory is the only known
candidate for a theory of quantum gravity, it is natural to ask if
string theory can provide a framework for flavor and CP conserving
gravity mediation. In compactified string theory, moduli (including
the dilaton) which determine the 4-dimensional (4D) gauge and Yukawa
couplings are the most plausible candidate for a messenger of SUSY
breaking, giving a gravity mediated contribution to gaugino and
sfermion masses \cite{brignole}. Then, constraints from low energy
flavor and/or CP violations imply that the dominant messenger
modulus should have flavor universal and CP conserving coupling to
the minimal supersymmetric standard model (MSSM) chiral matters. As
the mechanism of moduli stabilization determines which modulus is
the dominant messenger, this in turn leads to a nontrivial
constraint on the possible moduli stabilization scheme.

Moduli mediated SUSY breaking and its phenomenological consequences
have been studied before while regarding the moduli $F$ components
as a generic  background without specifying the underlying
stabilization scheme \cite{brignole}. It has been noticed that a
particular form of mediation dominated by the heterotic string
dilaton gives universal and CP conserving soft terms at string tree
level. If such dilaton domination can be realized while keeping the
quantum correction to the K\"ahler potential small enough, the
resulting soft terms would satisfy the constraints from flavor and
CP violation with sparticle masses in sub-TeV range
\cite{louis.nir}.

Recent progress of flux compactification suggests that string flux
might play key roles to achieve a phenomenologically viable string
vacuum state \cite{flux.review}. Flux can stabilize moduli while
producing a huge landscape of vacua which contains a de Sitter
vacuum with nearly vanishing cosmological constant.  Flux
compactification can also provide a SUSY breaking scheme in which
soft terms preserve flavor and CP at leading order in the string
coupling $g_{st}$ or the slope parameter $\alpha^\prime$
\cite{ibanez,choi,conlon2}. In this SUSY breaking scheme, only a
particular modulus which is unfixed by flux and whose vacuum
expectation value controls the $g_{st}$ or $\alpha^\prime$ expansion
of 4D effective action can be an important messenger of SUSY
breaking. The couplings of such messenger modulus to chiral matter
fields are naturally flavor universal and CP conserving at leading
order since the perturbative expansion is controlled by the
messenger modulus itself.

While providing a good starting point, this scheme does not assure
yet the absence of dangerous flavor or CP violation even when all
other mediations in the model preserve flavor and CP. There can be
higher order correction to the messenger modulus-matter couplings in
the K\"ahler potential, which is expected to be flavor non-universal
in general \cite{louis.nir}. Then the modulus mediation itself
associated with such higher order term might lead to a flavor or CP
violation exceeding the current experimental bound. In this paper,
we first discuss some features of flux compactification leading to a
SUSY breaking scheme which preserves flavor and CP at leading order
in the perturbative expansion controlled by the messenger modulus,
and then examine the constraints on the scheme coming from flavor
and/or CP violation induced by higher order K\"ahler potential.  It
is found that all phenomenological constraints can be satisfied even
for generic form of higher order K\"ahler potential and sparticle
spectra in the sub-TeV range, under plausible assumptions on the
size of higher order correction and flavor mixing angles.
This implies for instance that  mirage mediation
\cite{choi,cfnop,choijeongokumura,mirage1} involving such modulus
mediation together with an anomaly mediation of comparable size and
also the modulus-dominated mediation
\cite{ibanez,conlon,perturbative} realized in flux compactification
can be free from the SUSY flavor and CP problems, while giving
gaugino and sfermion masses in the sub-TeV range. Same statement
applies also to the axionic or deflected mirage mediation
\cite{deflectedmirage} in which gauge mediation of comparable size
is added to mirage mediation.

The organization of this paper is as follows. In section 2, we
discuss the relevant features of flux compactification leading to
the SUSY breaking scheme under consideration. In section 3, we
examine the structure of soft terms induced by higher order K\"ahler
potential together with the constraints from various flavor and/or
CP violating observables. Section 4 is the conclusion.

\section{relevant features of flux compactification}

\subsection{Moduli mass hierarchy}

In this paper, we will be focusing on flux compactification which
can realize the weak scale SUSY together with the high unification
scale\footnote{In fact, our analysis of flavor and CP constraints in
sec. 3 applies also to the intermediate string scale scenario
proposed in \cite{conlon}.} $M_{GUT}\sim 2\times 10^{16}$ GeV. In
such compactification, both the string scale $M_{st}$ and the
compactification scale $M_{KK}$ are comparable to the 4D Planck
scale $M_{Pl}\approx 2.4\times 10^{18}$ GeV or $M_{GUT}$. This
results in a big mass hierarchy between the heavy moduli $U$
stabilized by flux and the light moduli $T$ unfixed by flux.
In this subsection, we briefly discuss this moduli mass hierarchy,
while ignoring the little mass hierarchies of ${\cal O}(10-10^2)$
between $M_{Pl}$, $M_{st}$, and $M_{KK}$, i.e. while regarding
\bea
M_{st}\sim M_{KK}\sim M_{Pl}.
\eea

If one introduces  nonzero flux over a cycle ${\cal C}$ in compact
internal space, the modulus parameterizing the size of ${\cal C}$ is
stabilized generically with a SUSY preserving mass $m_U$ comparable
to $M_{st}$ \cite{flux.review}. In the language of 4D effective
theory, one finds \bea m_U\,\sim \,\left\langle \frac{\partial^2
W_{\rm flux}} {\partial U^2}\right\rangle\,\sim M_{st}, \eea where
$W_{\rm flux}$ is the flux-induced superpotential. Most string
compactifications allow the NS or RR 3-form fluxes over the 3-cycles
of internal space, which would stabilize all complex structure
moduli. Depending upon the model, string dilaton or K\"ahler moduli
might be stabilized also by flux. For instance, in type IIB
compactification, the dilaton can be stabilized by RR 3-form flux
\cite{gkp}.
It has been noticed that K\"ahler moduli in heterotic
compactification might be stabilized by intrinsic torsion flux
\cite{lukas}, suggesting the possibility that all complex structure
and K\"ahler moduli in heterotic compactifications are stabilized by
nonzero NS and torsion fluxes.

In many flux compactifications, there remains a modulus $T$ which
can {\it not} be fixed by flux.
One example of such modulus is the volume modulus in type IIB flux
compactification. The dilaton in heterotic string compactification
can be another example. Eventually, this modulus should be
stabilized by other means, e.g. a nonperturbative dynamics
\cite{kklt}. It is expected that the resulting modulus mass $m_T$ is
tied to the scale of SUSY breaking, and thus \bea m_T \,\sim\,
m_{3/2} \eea up to a little hierarchy of ${\cal O}(10-10^2)$.

In order to realize the weak scale SUSY, the gravitino mass
$m_{3/2}$ is required to be smaller than $M_{Pl}$ by many orders of
magnitudes. For $M_{st}\sim M_{Pl}$,  this is nontrivial to be
achieved in flux compactification as generic flux configuration
yields $\langle W_{\rm flux}\rangle={\cal O}(1)$ (in the unit with
$M_{Pl}=1$) due to the quantization of flux. On the other hand, if
SUSY is broken by a nonperturbative dynamics such as gaugino
condensation \cite{gauginocondensation}, or a warped dynamics
\cite{kachru}, the resulting SUSY breaking scale is hierarchically
lower than $M_{Pl}$: \bea \label{susyscale} M_{\rm SUSY}\sim
e^{-A}M_{Pl}, \eea where $e^{-A}$ is an exponentially small
nonperturbative or warp factor. In 4D effective theory, the vacuum
energy density at leading order is given by \bea V_{\rm vac}=M_{\rm
SUSY}^4-3m_{3/2}^2M_{Pl}^2, \eea where $m_{3/2}/M_{Pl}\sim \langle
W_{\rm flux}\rangle$. As a result, in nonperturbative or warped SUSY
breaking scenario, only a particular class of flux vacua with an
exponentially small vacuum value of the flux-induced superpotential,
i.e. \bea \langle W_{\rm flux}\rangle\sim e^{-2A}, \eea can have a
(nearly) vanishing cosmological constant.

With the above observation, one can make the following assumptions
to achieve a phenomenologically viable vacuum state with weak scale
SUSY: (i) the underlying compactification involves a large number
$N\gg 1$ of cycles each of which can carry a quantized flux in the
range $[-L,L]$ for $L\gg 1$, which would allow a huge number of
different flux configurations of ${\cal O}(L^N)$, (ii) such flux
configurations provide a fine discretum of $\langle W_{\rm
flux}\rangle$ varying from ${\cal O}(1)$ to a nearly vanishing
value, (iii) SUSY is broken by nonperturbative or warped dynamics,
yielding an exponentially small $M_{\rm SUSY}/M_{Pl}\sim
e^{-A}\sim\, 10^{-6}-10^{-7}.$
 To
be able to tune the vacuum energy density to the observed value
$\sim (3\times 10^{-12} {\rm GeV})^4$, the spacing between different
values of $\langle W_{\rm flux}\rangle$ should be as small as \bea
\delta\langle W_{\rm flux}\rangle\lesssim
\left(\frac{M_{Pl}}{m_{3/2}}\right) \left( \frac{3\times
10^{-12}{\rm GeV}}{M_{Pl}}\right)^4 \sim 10^{-104}. \eea Such
extremely fine spacing might be achieved in flux compactification
with $N\sim L={\cal O}(100)$ as in the case of flux energy density
discussed in \cite{busso}.

Under the assumptions specified above, the fine tuning for vanishing
cosmological constant selects a particular class of flux vacua with
$\langle W_{\rm flux}\rangle \sim e^{-2A}$. For such vacua, still
the moduli mass $m_U\sim\langle
\partial^2 W_{\rm flux}/\partial U^2\rangle$ is generically of order
unity due to the flux quantization. This results in a big moduli
mass hierarchy:
\bea
\label{masshierarchy}
\frac{m_T}{m_U}\sim
\frac{\langle W_{\rm flux}\rangle}{\langle
\partial^2 W_{\rm flux}/\partial U^2\rangle}
\sim \frac{m_{3/2}}{M_{Pl}}\sim e^{-2A},
\eea
where again the little hierarchy factors of ${\cal O}(10-10^2)$ are ignored.
It should be stressed that this moduli mass hierarchy is an outcome of the fine
tuning of the cosmological constant and the assumed hierarchy (\ref{susyscale})
between the SUSY breaking scale and the Planck scale.

Generically, both the heavy moduli $U$ and the light modulus $T$
couple to SUSY breaking sector, thereby develop nonzero
$F$-components. However, regardless of the details of SUSY breaking,
the $F$-component of the flux stabilized $U$ is given by \bea
F^U\,\sim\, \frac{m_{3/2}^2}{m_U}, \eea which is negligibly small
for $m_U$ comparable to the string or GUT scale. (Note that moduli
are normalized to be dimensionless, so their $F$ components have a
mass dimension one.) On the other hand, the light modulus $T$ can
develop a sizable $F^T$, e.g. \bea F^T \,\sim\, m_{3/2} \quad
\mbox{or}\quad \frac{m_{3/2}}{\ln(M_{Pl}/m_{3/2})}, \eea therefore
can be an important messenger of SUSY breaking
\cite{choi,conlon,perturbative}.

\subsection{4D effective action expanded in the inverse powers of the messenger modulus}

Quite often, the messenger modulus $T$ which is unfixed by flux has
the following features: (a) $1/{\rm Re}(T)$ is proportional to
certain powers of the string coupling $g_{st}$ or the inverse of the
compactification radius (in the unit with $\alpha^\prime=1$), thus
its vacuum expectation value controls the $g_{st}$ or
$\alpha^\prime$ expansion of the 4D action, (b) ${\rm Im}(T)$ is an
axion whose non-linear PQ symmetry
\bea
\label{PQ}
U(1)_T: \,\,{\rm Im}(T)\rightarrow {\rm Im}(T)+\mbox{constant}
\eea
is respected at any finite order in the $g_{st}$ and $\alpha^\prime$ expansion.
As a concrete example of such messenger modulus, one might consider the
volume modulus and its RR axion partner in type IIB flux compactification or the
dilaton-axion in heterotic flux compactification.

For such messenger modulus $T$, the couplings of moduli to the
visible gauge and matter fields are given by \bea \label{superspace}
&&\qquad\int d^4\theta \,\,Y_{I\bar{J}}(T+T^*,U,U^*)Q^I Q^{J*} \nonumber \\
&+&\left( \int d^2\theta \left[\frac{1}{4}f_a(T,U)
W^{a\alpha}W^a_\alpha + \frac{1}{6}\lambda_{IJK}(U)Q^I Q^J Q^K
\right]+{\rm h.c.}\right), \eea where $W^{a\alpha}$ and $Q^I$ denote
the visible gauge and matter superfields, respectively. Here the
matter kinetic function $Y_{I\bar{J}}$ is given by \bea
Y_{I\bar{J}}=e^{-K_0/3}Z_{I\bar{J}} \eea for the K\"ahler potential
\bea K = K_0+Z_{I\bar{J}} Q^I Q^{J*}, \eea where $K_0$ is the moduli
K\"ahler potential and $Z_{I\bar{J}}$ are the matter K\"ahler
metric. Expanding the 4D action in powers of $g_{st}$ or
$\alpha^\prime$ while preserving the non-linear PQ symmetry
$U(1)_T$, the matter and gauge kinetic functions can be written as
\bea Y_{I\bar{J}} &=& (T+T^*)^{n_{I\bar{J}}}\Gamma_{I\bar{J}}(U,U^*)
\left(1-\frac{\Delta_{I\bar{J}}(U,U^*)}{[8\pi^2(T+T^*)]^{k_{I\bar{J}}}}+...\right),
\nonumber \\
f_a&=& k_aT+\frac{1}{8\pi^2}\Delta_a(U),
\nonumber
\eea
where $n_{I\bar{J}},k_{I\bar{J}}$, and $k_a$ are all rational numbers.
The successful unification of the MSSM gauge couplings at
$M_{GUT}\sim 2\times 10^{16}$ GeV suggests that $k_a$ are universal
for the MSSM gauge group. In the following, we take the
normalization of $T$ for which $k_a=1$, and thus
\bea
\langle {\rm Re}(T)\rangle \simeq \langle {\rm Re}(f_a)\rangle \simeq
\frac{1}{g_{GUT}^2}.
\eea

As was noticed before \cite{choi,conlon1,Choi:2006za}, if the MSSM
chiral matters with same gauge charge originate from branes with
same world volume dimension, the matter modular weights
$n_{I\bar{J}}$ are automatically flavor universal (see Appendix A
for a more discussion of matter modular weights): \bea
n_{I\bar{J}}=\mbox{flavor universal}\,\, n_I. \eea Also, in view of
that $T$ determines the 4D gauge coupling, it is expected that the
messenger modulus expansion of 4D action is controlled by \bea
\frac{1}{8\pi^2(T+T^*)}\sim \frac{\alpha_{GUT}}{4\pi}, \eea
and thus
\bea
k_{I\bar{J}}=1, \quad \Delta_{I\bar{J}}={\cal O}(1),
\quad \Delta_a={\cal O}(1).
\eea
In the following, we will assume this feature of the messenger modulus expansion,
and examine its phenomenological consequences. Note that the non-linear PQ symmetry
$U(1)_T$ and the holomorphicity assure that $\lambda_{IJK}$ are
independent of $T$.

At leading order in the messenger modulus expansion, the non-linear PQ symmetry
$U(1)_T$ and the flavor universality of matter modular weights $n_I$  assure that
\bea
&&\frac{\partial}{\partial T}\ln(Y_{I\bar{J}}) =\frac{n_I}{T+T^*}=
\mbox{real and flavor universal},
\nonumber \\
&& \frac{\partial}{\partial T}\ln (\lambda_{IJK})=0,
\nonumber \\
&& \frac{\partial}{\partial T}\ln({\rm Re}(f_a))=\frac{k_ag_a^2}{2}
=\mbox{real}, \eea with which the $T$-mediated SUSY breaking
preserves flavor and CP \cite{choi,conlon,Choi:1993yd}.
 On the other
hand, $\frac{\partial}{\partial U}\ln(Y_{I\bar{J}})$ and
$\frac{\partial}{\partial U}\ln (\lambda_{IJK})$ are flavor
non-universal and complex, so the $U$-mediated SUSY breaking
violates flavor and CP in general. However, as we will see shortly,
 \bea
\quad F^U \sim \frac{m_{3/2}^2}{m_U}\sim \frac{m_{3/2}^2}{M_{Pl}}
\eea regardless of the details of SUSY breaking, and thus the moduli
mass hierarchy (\ref{masshierarchy}) assures that the $U$-mediated
SUSY breaking is absolutely negligible. Still there might be a
dangerous CP violation associated with the phase of Higgs $\mu$ and
$B$ parameters. Even for this, the non-linear PQ symmetry $U(1)_T$
is useful as it allows the relative phase between $F^T$ and
$m_{3/2}$ to be rotated away. With real $F^T/m_{3/2}$, if $\mu$ is
generated dominantly by the Chun-Kim-Nilles mechanism \cite{chun},
or by the Giudice-Masiero mechanism \cite{giudice}, or by a singlet
vacuum value as in the next to minimal supersymmetric standard
model, the resulting Higgs mass parameters preserve CP
\cite{choijeongokumura}.

One can now integrate out the heavy moduli $U$ to derive the
effective action of the visible fields and the light messenger
modulus $T$. Let us start with the full 4D action which is
generically given by
\bea
\label{fullaction}
\int d^4\theta\,
CC^*\,\Omega(U,U^*,\Phi,\Phi^*)+\left[\int d^2\theta C^3\Big(W_{\rm flux}(U)
+ \tilde W(U,\Phi)\Big)+{\rm h.c.}\right],
\eea
where $C$ is the chiral compensator superfield and $\Phi$ stands for all light
superfields including the visible gauge and matter fields as well as
the light modulus $T$. Here  $W_{\rm flux}$ is the flux-induced
superpotential depending only on $U$, and $\tilde W$ denotes the
other part of superpotential which might include a $U(1)_T$ breaking
non-perturbative term, e.g.
\bea
\tilde W =A(U)e^{-aT}+\frac{1}{6}\lambda_{IJK}(U)Q^I Q^J Q^K.
\eea

To integrate out $U$, we note that flux quantization implies \bea
\label{fluxinducedmass} M_U\equiv \frac{\partial^2 W_{\rm
flux}(U=U_0)}{\partial U^2} \,\sim\, M_{st}, \eea and the fine
tuning of the cosmological constant in the presence of
non-perturbative or warped SUSY breaking requires \bea W_{\rm
flux}(U=U_0)\,\sim\, e^{-2A}, \eea where $e^{-A}=M_{SUSY}/M_{Pl}$ is
an exponentially small non-perturbative or warp factor and $U_0$ is
the globally supersymmetric stationary point of the flux-induced
superpotential: \bea \frac{\partial W_{\rm flux}(U=U_0)}{\partial
U}=0. \eea Apparently the physical moduli mass $m_U$ is dominated by
the globally supersymmetric mass $M_U$ in the limit when $M_U\gg
m_{3/2}$, and $U_0$ and $M_U$ are independent of light superfields.

The heavy moduli $U$ can be integrated out by replacing $U$ in the
action (\ref{fullaction}) with the solution of the  following
superfield equation of motion: \bea \frac{1}{4}\bar{{\cal
D}}^2\left(CC^*\frac{\partial \Omega}{\partial U}\right) +
C^3\frac{\partial W}{\partial U}=0,\eea where $\bar{\cal
D}^2=\bar{\cal D}^{\dot{\alpha}}\bar{\cal D}_{\dot{\alpha}}$ denotes
the supercovariant derivative, $W=W_{\rm flux}+\tilde W$ and all
light fields $\Phi$ and also the compensator $C$ are considered to
be generic background superfields. In the limit with
$m_{3/2}/M_U\sim e^{-2A}\ll 1$, the solution can be expanded in
powers of $\bar{{\cal D}}^2/M_U$ and $\tilde W/M_U$ both of which
are of the order of $m_{3/2}/M_U$. Note that ${\partial^n \tilde
W}/{\partial U^n} \sim m_{3/2}$ for arbitrary $n\geq 0$ if the mass
scale of the visible sector, e.g. the weak scale, is determined by
SUSY breaking. In the perturbative expansion in powers of
$\bar{{\cal D}}^2/M_U$ and $\tilde W/M_U$, the solution is given by
\bea U=U_0-\frac{1}{M_U}\left[ \frac{1}{4}\bar{{\cal D}}^2
\left(\frac{C^*}{C^2}\frac{\partial\Omega(U_0,U_0^*,\Phi,\Phi^*)}{\partial
U}\right) +\frac{\partial \tilde W(U_0,\Phi)}{\partial U}\right]
+..., \eea where $M_U$ is given in (\ref{fluxinducedmass}), and the
ellipsis denotes higher order terms. One immediate consequence of
this superfield solution is \bea F^U \sim \frac{m_{3/2}}{M_U}F^\Phi
\sim \frac{m_{3/2}^2}{M_U}, \eea which assures that $F^U$ is
negligibly small compared to $F^\Phi\sim m_{3/2}$ when $M_U\sim
M_{st}$.

It is now obvious that, upon ignoring the small corrections
suppressed by $m_{3/2}/M_U$, the low energy effective lagrangian can
be obtained by replacing $U$ in (\ref{fullaction}) with $U_0$. After
this, one can make a proper redefinition of $Q^I$ under which \bea
\Gamma_{I\bar{J}}(U_0,U_0^*)\rightarrow \delta_{I\bar{J}},\quad
\Delta_{I\bar{J}}(U_0,U_0^*)\rightarrow \Delta_I\delta_{I\bar{J}}.
\eea After such field redefinition, the effective couplings of the
messenger modulus $T$ to the visible gauge and matter fields are
given by \bea \label{effectiveaction} \int
d^4\theta\,Y_IQ^IQ^{I*}+\left(\int d^2\theta
\left[\frac{1}{4}f_aW^aW^a + \frac{1}{6}\lambda_{IJK}Q^I Q^J
Q^K\right]+{\rm h.c.}\right)+{\cal O}
\left(\frac{m_{3/2}}{M_U}\right), \eea where \bea \label{effective}
Y_I &=& (T+T^*)^{n_I}\left(1-\frac{\Delta_I}{8\pi^2(T+T^*)}\right),
\nonumber \\
f_a&=& k_aT + \frac{1}{8\pi^2}\Delta_a, \eea where $\Delta_I$ and
$\Delta_a$ are constants of order unity, and $\lambda_{IJK}$ are
constants with which the canonically normalized  Yukawa couplings
are determined as \bea y_{IJK}=\frac{\lambda_{IJK}}{\sqrt{Y_I Y_J
Y_K}} \simeq \frac{\lambda_{IJK}}{(T+T^*)^{(n_I+n_J+n_K)/2}}. \eea

The soft SUSY-breaking terms of canonically normalized sfermion
fields $\tilde{Q}^I$ can be written as \bea {\cal L}_{\rm
soft}=-\frac{1}{2}m_I^2|\tilde{Q}^I|^2
-\frac{1}{6}A_{IJK}y_{IJK}\tilde{Q}^I\tilde{Q}^J\tilde{Q}^K+{\rm
h.c.}, \eea which include the modulus mediated contribution
\cite{brignole} at $M_{GUT}$ as \bea
m_I^2&=&-F^TF^{\bar{T}}\partial_T\partial_{\bar{T}}\ln
\left(Y_I\right)+... \nonumber \\
&=& \left(n_I+\frac{g_{GUT}^2}{8\pi^2}\Delta_I\right)M_0^2+...,
\nonumber \\
A_{IJK}&=&
-F^T\partial_T\ln\left(\frac{\lambda_{IJK}}{Y_IY_JY_K}\right)+...\nonumber
\\
&=& \left[\left(n_I+n_J+n_K\right)+\frac{g_{GUT}^2}{16\pi^2}
\left(\Delta_I+\Delta_J+\Delta_K\right)\right]M_0+..., \eea where
\bea M_0=\frac{F^T}{T+T^*} \eea corresponds to the modulus mediated
contribution to the gaugino mass at $M_{GUT}$, and the ellipses
stand for the contribution from other mediation in the model.

The matter modular weights $n_I$ are typically flavor universal,
however there is no a priori reason for the higher order
coefficients $\Delta_I$ to be flavor universal also. Even when
$\Delta_I$ are flavor non-universal, there would not be any
dangerous flavor or CP violation if sfermion masses are much heavier
than 1 TeV, which actually happens for instance in the scheme
proposed in \cite{matteruplifting}. In this paper, we are concerned
with the possibility that modulus mediation including higher order
effects satisfies the flavor and CP constraints with sparticle
spectra in the sub-TeV range. To see this, we will examine in the
next section the constraints on $\Delta_I$ imposed by low energy
flavor and/or CP violating observables under the assumption that
they are the dominant origin of non-minimal flavor or CP violation.

\section{constraints from flavor and/or cp violation}

Let us first set up the notation. We start with the field basis for
which the matter kinetic functions are diagonal as in the effective
action (\ref{effectiveaction}). The MSSM matters and their $N=1$
superspace kinetic functions are denoted as \bea Q^I&=&\{q_i,
u^c_i,d^c_i,l_i,e^c_i\},
\nonumber \\
Y_I&=&\{Y^q_i,Y^u_i,Y^d_i,Y^l_i,Y^e_i\}, \eea where $q_i$
($i=1,2,3$) are the $SU(2)_W$ doublet quarks, $u^c_i$ and $d^c_i$
are the $SU(2)_W$ singlet anti-quarks, $l_i$ are the $SU(2)_W$
doublet leptons, $e^c_i$ are the $SU(2)_W$ singlet leptons, and the
matter kinetic functions include higher order correction as \bea
Y^\phi_i=(T+T^*)^{n_\phi}\left(1-\frac{\Delta^\phi_i}{8\pi^2(T+T^*)}\right)\qquad
(\phi=q,u,d,l,e). \eea Here we are interested in the flavor or CP
violations associated with  $\Delta^\phi_i-\Delta^\phi_j\neq 0$ for
$i\neq j$ as the higher order correction to the gauge kinetic
function, i.e. $\Delta_a$ of (\ref{effective}), obviously preserves
flavor, and also does not give any CP violation.

Yukawa couplings and soft SUSY breaking terms of the canonically normalized MSSM
matters at the {\it weak scale} are parameterized as
\bea
{\cal L}_{\rm Yukawa} &=& y^u_{ij}H_uq_iu^c_j+y^d_{ij} H_d q_id^c_j +
y^e_{ij}H_dl_ie^c_j+\kappa^\nu_{ij}H_ul_iH_ul_j + {\rm h.c.},
\nonumber \\
{\cal
L}_{\rm soft}&=&-\left(\,
A^u_{ij}y^u_{ij}H_u\tilde{q}_i\tilde{u}^c_j
+ A^d_{ij}y^d_{ij}\tilde{q}_iH_d\tilde{d}^c_j
+ A^e_{ij}y^e_{ij}H_d\tilde{l}_i\tilde{e}^c_j + {\rm h.c.} \,\right)
\nonumber \\
&& -\left(\,m^{2(\tilde{q})}_{{i}j}\tilde{q}^*_i \tilde{q}_j
+m^{2(\tilde{u})}_{i{j}}\tilde{u}^c_i \tilde{u}_j^{c*}
+m^{2(\tilde{d})}_{i{j}}\tilde{d}^c_i \tilde{d}_j^{c*}
+m^{2(\tilde{l})}_{{i}j}\tilde{l}^*_i \tilde{l}_j
+m^{2(\tilde{e})}_{i{j}}\tilde{e}^c_i \tilde{e}_j^{c*}\,\right),
\eea where we include the $D=5$ operator for neutrino masses in
${\cal L}_{\rm Yukawa}$. Soft parameters can be decomposed as \bea
m^{2(\tilde{\phi})}_{i{j}}&=&m^{2(\tilde{\phi})}_0\delta_{i{j}}+\Delta
m^{2(\tilde{\phi})}_{i{j}} \qquad
(\tilde{\phi}=\tilde{q},\tilde{u},\tilde{d},\tilde{l},\tilde{e}),
\nonumber\\
A^\psi_{ij}&=&A^\psi_0+\Delta A^\psi_{ij} \qquad(\psi=u,d,e), \eea
where $m^{2(\tilde{\phi})}_0$ and $A^\psi_0$ stand for flavor
universal sfermion masses and $A$-parameters, respectively, while
$\Delta m^{2(\tilde{\phi})}_{i{j}}$ and  $\Delta A^\psi_{ij}$
represent flavor non-universal part. Depending upon the underlying
SUSY breaking scheme, $m^{2(\tilde{\phi})}_0$ and $A^\psi_0$ might
receive contributions from various sources, e.g. modulus mediation,
gauge mediation, anomaly mediation, renormalization group effect,
e.t.c., whose relative importance will depend on the details of the
model. Here we do not specify the full origin of the flavor
universal $m_0^{2(\tilde{\phi})}$ and $A^\psi_0$, however the flavor
non-universal part is assumed to be dominated by the modulus
mediated contribution associated with non-universal $\Delta^\phi_i$:
\bea \Delta m^{2(\tilde{\phi})}_{i{j}}&\simeq&
-\left[F^TF^{\bar{T}}\partial_T\partial_{\bar{T}}
\ln\left(1-\frac{\Delta^\phi_i}{8\pi^2(T+T^*)}\right)\right]\delta_{i{j}}
\nonumber \\
&\simeq& \frac{g_{GUT}^2}{8\pi^2}\Delta_i^\phi M_0^2\delta_{i{j}}
\qquad (\phi=q,u,d,l,e), \nonumber \\\Delta A^u_{ij}&\simeq&
F^T\partial_T\ln\left(1-\frac{\Delta^q_i}{8\pi^2(T+T^*)}\right)
\left(1-\frac{\Delta^u_j}{8\pi^2(T+T^*)}\right)
\nonumber \\
&\simeq& \frac{g_{GUT}^2}{16\pi^2}(\Delta_i^q+\Delta_j^u)M_0,
\nonumber \\
\Delta A^d_{ij}&\simeq&
F^T\partial_T\ln\left(1-\frac{\Delta^q_i}{8\pi^2(T+T^*)}\right)
\left(1-\frac{\Delta^d_j}{8\pi^2(T+T^*)}\right)
\nonumber \\
&\simeq& \frac{g_{GUT}^2}{16\pi^2}(\Delta_i^q+\Delta_j^d)M_0,
\nonumber \\
\Delta A^e_{ij}&\simeq&
F^T\partial_T\ln\left(1-\frac{\Delta^l_i}{8\pi^2(T+T^*)}\right)
\left(1-\frac{\Delta^e_j}{8\pi^2(T+T^*)}\right)
\nonumber \\
&\simeq& \frac{g_{GUT}^2}{16\pi^2}(\Delta_i^l+\Delta_j^e)M_0, \eea
where \bea M_0=\frac{F^T}{T+T^*} \eea corresponds to the modulus
mediated contribution to the gaugino mass at $M_{GUT}$. In fact,
there are renormalization group (RG) corrections to the above
non-universal part of soft parameters at the weak scale, which are
mostly due to the 3rd generation Yukawa couplings. However such RG
corrections can be safely ignored here as all the meaningful flavor
and CP constraints on the modulus mediated SUSY breaking under
consideration come from the first two generations for which the
Yukawa induced RG corrections are negligibly small.

To examine the flavor and/or CP violating observables induced by
$\Delta m^{2(\tilde{\phi})}_{i{j}}$ and $\Delta A^\psi_{ij}$, it is
convenient to use the super-CKM basis in which the quark and lepton
mass matrices are diagonal \cite{Dugan:1984qf}. Starting from the Yukawa coupling
matrices $y^\psi_{ij}$ ($\psi=u,d,e$) defined in the field basis for
which the matter kinetic functions are diagonal, the super-CKM basis
can be achieved by the unitary rotations of the matter superfields
under which the Yukawa matrices become real and diagonal:
\bea
\label{diagonalization}
(V^{\psi}_L)^T y^\psi V^{\psi}_R &=& {\rm Diag}(
\hat y^\psi_1, \hat y^\psi_2, \hat y^\psi_3 ),
\nonumber \\
(V^{\nu}_L)^T \kappa^{\nu} V^{\nu}_L &=& {\rm Diag}
(\hat\kappa^\nu_1, \hat\kappa^\nu_2, \hat\kappa^\nu_3 ),
\eea
where $V^\psi_{L,R}$ and $V^{\nu}_L$ are unitary matrices.

In supersymmetric limit, flavor and/or CP violations are all
described by  the CKM and PMNS mixing matrices given by \bea V_{\rm
CKM} = V^{u\dagger}_L V^{d}_L, \quad V_{\rm PMNS} = V^{e\dagger}_L
V^{\nu}_L. \eea However, in the presence of soft SUSY breaking
terms, there can be further flavor and/or CP violations induced by
non-universal $\Delta m^{2(\tilde{\phi})}_{i{j}}$ and $\Delta
A^\psi_{ij}$. Most of those non-minimal flavor violations can be
described by the following mass-insertion parameters with $i\neq j$
\cite{Hall:1985dx, Kobayashi:2000br}:
\bea \left(\delta^d_{LL}\right)_{i{j}}&=&\frac{(V_L^{d\dagger}
\Delta m^{2(\tilde{q})}V_L^{d})_{i{j}}}{m^2_{\tilde{q}}}\simeq
\frac{g_{GUT}^2}{8\pi^2}\frac{M_0^2}{m^2_{\tilde{q}}}\left(\Delta_{LL}^d\right)_{i{j}},
\nonumber \\
\left(\delta^d_{RR}\right)_{i{j}}&=&\frac{(V_R^{dT} \Delta
m^{2(\tilde{d})}V_R^{d*})_{i{j}}}{m^2_{\tilde{q}}}\simeq
\frac{g_{GUT}^2}{8\pi^2}\frac{M_0^2}{m^2_{\tilde{q}}}\left(\Delta_{RR}^d\right)_{i{j}},
\nonumber \\
\left(\delta^d_{LR}\right)_{ij}&=&\frac{(V_L^{d T}\Delta{\cal
A}^dV^{d}_R)_{ij}\langle H_d\rangle}{m^2_{\tilde{q}}}\simeq
\frac{g_{GUT}^2}{16\pi^2}\frac{M_0^2}{m^2_{\tilde{q}}}\left(
\left(\Delta_{LL}^d\right)_{ij}\frac{m^d_j}{M_0}+\frac{m^d_i}{M_0}
\left(\Delta_{RR}^d\right)_{ij}\right),
\nonumber \\
\left(\delta^e_{LL}\right)_{i{j}}&=&\frac{(V_L^{e\dagger} \Delta
m^{2(\tilde{l})}V_L^{e})_{i{j}}}{m^2_{\tilde{l}}}\simeq
\frac{g_{GUT}^2}{8\pi^2}\frac{M_0^2}{m^2_{\tilde{l}}}\left(\Delta_{LL}^e\right)_{i{j}},
\nonumber \\
\left(\delta^e_{RR}\right)_{i{j}}&=&\frac{(V_R^{eT} \Delta
m^{2(\tilde{e})}V_R^{e*})_{i{j}}}{m^2_{\tilde{l}}}\simeq
\frac{g_{GUT}^2}{8\pi^2}\frac{M_0^2}{m^2_{\tilde{l}}}\left(\Delta_{RR}^e\right)_{i{j}},
\nonumber \\
\left(\delta^e_{LR}\right)_{ij}&=& \frac{(V_L^{e T}\Delta{\cal
A}^eV^{e}_R)_{ij}\langle H_d\rangle}{m^2_{\tilde{l}}}\simeq
\frac{g_{GUT}^2}{16\pi^2}\frac{M_0^2}{m^2_{\tilde{l}}}\left(
\left(\Delta_{LL}^e\right)_{ij}\frac{m^e_j}{M_0}+\frac{m^e_i}{M_0}
\left(\Delta_{RR}^e\right)_{ij}\right), \eea
 where $m_{\tilde{q}}$ and $m_{\tilde{l}}$ denote the average squark
 and slepton masses,  $m^d_i$
and $m^e_i$ ($i=1,2,3$) are the down-type quark and charged lepton
mass eigenvalues, and  \bea
\left(\Delta^{d,e}_{LL}\right)_{ij}&=&
\sum_k(V_L^{d,e})^*_{ki}(V_L^{d,e})_{kj}\Delta^{q,l}_k,
\nonumber \\
\left(\Delta^{d,e}_{RR}\right)_{ij}&=&
\sum_k(V_R^{d,e})_{ki}(V_R^{d,e})^*_{kj}\Delta^{d,e}_k,
\nonumber \\
\Delta{\cal A}^{d,e}_{ij}&=&y^{d,e}_{ij}\Delta A^{d,e}_{ij}.
\eea

According to our assumption that the messenger modulus expansion is
controlled by $1/8\pi^2{\rm Re}(T)$, all of the above mass-insertion
parameters are suppressed by a factor of ${\cal
O}(g_{GUT}^2/8\pi^2)$. In fact, the flavor changing mass-insertion
parameters with $i\neq j$ can be further suppressed by small mixing
angle in the unitary matrices $V_{L,R}^\psi$ ($\psi=u,d,e$). To see
this,  we note that the observed quark and charged lepton masses and
the CKM mixing angles suggest that the Yukawa couplings take the
form \bea \label{yukawa_pattern}y^u_{ij}\sim
\epsilon_i^q\epsilon_i^{u}, \quad y^d_{ij}\sim
\epsilon_i^q\epsilon_j^{d},\quad y^e_{ij}\sim
\epsilon^l_i\epsilon^{e}_j.\eea This form of Yukawa couplings can be
naturally obtained by assuming either the localization of matter
fields in extra dimension
\cite{Arkani-Hamed:1999dc,Ibanez:1986ka,Cremades:2002va} or a
spontaneously broken flavor symmetry \cite{Froggatt:1978nt}. In the
scheme utilizing localization, different flavors with the same gauge
charge are assumed to be localized at different positions in extra
dimension, and then the flavor parameters $\epsilon^\phi_i$
($\phi=q,u,d,l,e$) determined  by the wavefunction of matter fields
show hierarchical pattern.   Similar result can be obtained also in
the scheme which assumes a broken flavor symmetry under which
different flavors have different charges. In both schemes, the above
form of Yukawa couplings is maintained even after the kinetic terms
of matter fields are diagonalized. Note that neither localization
nor flavor symmetry does provide a further suppression of
$\Delta^\phi_i$ in the matter kinetic functions.

The Yukawa couplings of (\ref{yukawa_pattern}) give rise to the mass
hierarchy:\bea \label{mass}{m^u_i}/{m^u_j}\sim
|{\epsilon^q_i\epsilon^{u}_i}|/|{\epsilon^q_j\epsilon^{u}_j}|,\quad
{m^d_i}/{m^d_j}\sim
|{\epsilon^q_i\epsilon^{d}_i}|/|{\epsilon^q_j\epsilon^{d}_j}|,\quad
{m^e_i}/{m^e_j}\sim
|{\epsilon^l_i\epsilon^{e}_i}|/|{\epsilon^l_j\epsilon^{e}_j}|,\eea
and also  the mixing angle pattern for $i\leq j$: \bea
\label{mixing} \big(V_L^{u,d}\big)_{ij}&\sim&
\big(V_L^{u,d}\big)_{ji}\,\sim\,
{\epsilon_i^q}/{\epsilon_j^q},\qquad \big(V_L^e \big)_{ij}\,\sim\,
\big(V_L^e \big)_{ji}\,\sim\,
{\epsilon_i^l}/{\epsilon_j^l},\nonumber
\\
\big(V_R^{\psi}\big)_{ij}&\sim& \big(V_R^{\psi}\big)_{ji}\,\sim\,
{\epsilon_i^\psi}/{\epsilon_j^\psi}, \quad(\psi=u,d,e),\eea where we
have assumed the normal hierarchy structure: \bea
|\epsilon^\phi_1|\lesssim|\epsilon^\phi_2|\lesssim|\epsilon^\phi_3|.\eea
This pattern of mixing angles implies for instance \bea
\label{massratio} \left|
\big(V^{d}_L\big)_{12}\big(V_R^{d}\big)_{12}\right|\sim
m_d/m_s,\quad
\left|\big(V^{e}_L\big)_{12}\big(V_R^{e}\big)_{12}\right|\sim
m_e/m_\mu.
\eea

Using the mass hierarchy (\ref{mass}) and the mixing angle pattern
(\ref{mixing}) together with \bea \sum_kV^*_{ki}V_{kj}\Delta_k &=&
\delta_{ij}\Delta_1+V_{2i}^*V_{2j}(\Delta_2-\Delta_1)+V_{3i}^*V_{3j}(\Delta_3-\Delta_1)
\nonumber \\
&=&\delta_{ij}\Delta_2+ V_{1i}^*V_{1j}(\Delta_1-\Delta_2)
+V^*_{3i}V_{3j}(\Delta_3-\Delta_2),\eea  it is straightforward to
find  (for $i\neq j$)
 \bea \label{MId1}
(\delta^{d}_{LL})_{ij} &\sim&
\frac{g_{GUT}^2}{8\pi^2}\left(\frac{M^2_0}{m^2_{\tilde q}}\right)
(\Delta^{q}_j - \Delta^{q}_i)\left(V_L^d\right)_{ij},
\nonumber \\
(\delta^{d}_{RR})_{ij} &\sim&
\frac{g_{GUT}^2}{8\pi^2}\left(\frac{M^2_0}{m^2_{\tilde q}}\right)
(\Delta^{d}_j - \Delta^{d}_i)\left(V_R^d\right)_{ij},
\nonumber \\
(\delta^{e}_{LL})_{ij} &\sim&
\frac{g_{GUT}^2}{8\pi^2}\left(\frac{M^2_0}{m^2_{\tilde l}}\right)
(\Delta^{l}_j - \Delta^{l}_i)\left(V_L^e\right)_{ij},
\nonumber \\
(\delta^{e}_{RR})_{ij} &\sim&
\frac{g_{GUT}^2}{8\pi^2}\left(\frac{M^2_0}{m^2_{\tilde l}}\right)
(\Delta^e_j - \Delta^e_i)\left(V_R^e\right)_{ij},
\nonumber \\
(\delta_{LR}^{d,e})_{ij}&\sim&
\frac{m_i^{d,e}}{2M_0}(\delta_{RR}^{d,e})_{ij}
+\frac{m_j^{d,e}}{2M_0}(\delta_{LL}^{d,e})_{ij}.
\eea

Let us now consider the phenomenological constraints on the mass-insertion parameters.
For the quark sector, the most stringent
constraint comes from the CP violating  $K$-$\bar K$ mixing
parameter $\epsilon_K$. Requiring that the SUSY contribution to
$\epsilon_K$ should be less than the standard model value
\footnote{In view of that the CKM phase explains rather accurately
all the observed CP violating phenomena including those of the $B$
meson system,  one might require a stronger condition that the SUSY
contribution to $\epsilon_K$ should be less than about 10\% of the
standard model prediction. This would result in a factor of few
stronger bound than (\ref{epsilonkbound}), but does not change our
conclusion.}, while assuming the gluino mass $m_{\tilde{g}}\sim
m_{\tilde{q}}$, one finds \cite{Ciuchini:1998ix} \bea
\label{epsilonkbound} \sqrt{\left|\,{\rm
Im}\left[(\delta^d_{LL})_{12}(\delta^d_{RR})_{12}\right] \right|}
&\lesssim& 4 \times 10 ^{-4} \left(\frac{m_{\tilde q}}{1\,{\rm
TeV}}\right),
\nonumber \\
\sqrt{\left|\,{\rm Im}\left[(\delta^d_{LR,RL})^2_{12}\right]
\right|} &\lesssim& 8 \times 10 ^{-4} \left(\frac{m_{\tilde
q}}{1\,{\rm TeV}}\right).
\eea
For the mass-insertion parameters  of (\ref{MId1}), the second
bound is easily satisfied, while the first bound leads to  \bea
\frac{M_0^2}{m_{\tilde{q}}^2}\left(\frac{m_d}{m_s}\right)^{1/2}
\sqrt{\left|(\Delta^q_2-\Delta^q_1)(\Delta^d_2-\Delta^d_1)\sin\eta_d\right|}
&\lesssim& 7 \times 10 ^{-2} \left(\frac{m_{\tilde q}}{1\,{\rm
TeV}}\right), \eea where $\eta_d$ is a CP violating phase coming
from the unitary rotation matrices, and we have used the relation
$|(V_L^d)_{12}(V_R^d)_{12}|\sim m_d/m_s$.
Due to the renormalization group evolution, the squark mass $m_{\tilde{q}}$ at
the weak scale is typically bigger than the modulus mediated gaugino
mass $M_0$ at $M_{GUT}$. Then, with the help from the small mixing
angle $|(V^{d}_L)_{12}(V_R^{d})_{12}|\sim m_d/m_s \sim 1/20$ and
also an additional minor suppression  by $M_0^2/m_{\tilde{q}}^2\sim
1/3$, the above bound can be satisfied even when
$|\Delta^\phi_{1}-\Delta^\phi_2|\sim 1$, $|\sin\eta_d|\sim 1$, and
$m_{\tilde{q}}\sim 1$ TeV.

One might consider the $b \to s \gamma$ process to see if the higher
order matter kinetic functions give rise to a contribution exceeding
the current experimental bound. Requiring that the SUSY contribution
to the branching ratio of $b\to s\gamma$ should be less than $10^{-5}$,
again with $m_{\tilde{g}}\sim m_{\tilde{q}}$, one finds \cite{cjo7}
\bea
\left|(\delta^d_{LR})_{23}\right| &\lesssim& 5\times 10^{-3}
\left(\frac{m_{\tilde q}}{1\,{\rm TeV}}\right),\eea
which is well satisfied by the mass-insertion parameters estimated
in (\ref{MId1}).

One might consider also the atomic and neutron electric dipole
moments (EDMs) induced by the imaginary part of the diagonal LR
mass-insertion parameter \cite{Abel:2001vy}.
However, in our case those LR parameters are
given by \bea \label{edmsource} \left(\delta^{d}_{LR}\right)_{ii}&=&
\frac{(V_L^{d T}\Delta {\cal A}^{d}V^{d}_R)_{ii} \langle
H_d\rangle}{m^2_{\tilde{q}}}
\,\simeq\,
\frac{g_{GUT}^2}{16\pi^2}\frac{M_0^2}{m^2_{\tilde{q}}}\frac{m^{d}_i}{M_0}
\left[\left(\Delta_{LL}^d\right)_{ii}+\left(\Delta_{RR}^d\right)_{ii}\right],
\nonumber \\
\left(\delta^{e}_{LR}\right)_{ii}&=& \frac{(V_L^{e T}\Delta{\cal
A}^{e}V^{e}_R)_{ii}\langle H_d\rangle}{m^2_{\tilde{l}}}
\,\simeq\,
\frac{g_{GUT}^2}{16\pi^2}\frac{M_0^2}{m^2_{\tilde{l}}}\frac{m^{e}_i}{M_0}
\left[\left(\Delta_{LL}^e\right)_{ii}+\left(\Delta_{RR}^e\right)_{ii}\right],
\eea which are manifestly real. As a result, the atomic  and neutron
EDMs induced by higher order matter kinetic function are far below
the current experimental limits.

In fact, the most stringent constraints on the modulus mediated SUSY
breaking scheme  come from the $\mu\to e\gamma$ process. Requiring
that ${\rm Br}(\mu\to e\gamma)\leq 1.2\times10^{-11}$, while
assuming the Wino mass $m_{\tilde{W}}\sim m_{\tilde{l}}$ and the
Higgsino mass $\mu\sim 2m_{\tilde{l}}$, one finds \cite{Brignole:2004ah,
cjo7}
\bea
\left|(\delta^e_{LL})_{12} \right| &\lesssim& \frac{7\times
10^{-3}}{\tan\beta} \left(\frac{m_{\tilde l}}{300\,{\rm GeV}}
\right)^2,
\nonumber \\
\left|(\delta^e_{RR})_{12} \right| &\lesssim& \frac{2\times
10^{-2}}{\tan\beta} \left(\frac{m_{\tilde l}}{300\,{\rm GeV}}
\right)^2,
\nonumber \\
\left|(\delta^e_{LR,RL})_{12} \right| &\lesssim& 6\times 10^{-6}
\left(\frac{m_{\tilde l}}{300\,{\rm GeV}} \right).\eea
For the mass-insertion parameters given by (\ref{MId1}), the LR
bound is easily satisfied. On the other hand, the  LL and RR bounds
lead to  \bea \frac{M_0^2}{m^2_{\tilde{l}}}
\left|\big(V_L^e\big)_{12}\big(\Delta_1^l-\Delta_2^l\big)\right|
\lesssim \frac{1}{\tan\beta}\left(\frac{m_{\tilde l}}{300\,{\rm
GeV}} \right)^2,
\nonumber \\
\frac{M_0^2}{m^2_{\tilde{l}}}
\left|\big(V_R^e\big)_{12}\big(\Delta^e_1-\Delta_2^e\big)\right|
\lesssim \frac{3}{\tan\beta}\left(\frac{m_{\tilde l}}{300\,{\rm
GeV}} \right)^2.
\eea

It is reasonably expected that $M_0\sim m_{\tilde{l}}$, and also the
lepton mixing angles which affect $\mu\to e\gamma$ are related to
the $\mu$ to $e$ mass ratio as \bea
\left|\big(V_L^e\big)_{12}\big(V_R^e\big)_{12}\big)\right|\,\sim\,
\frac{m_e}{m_\mu}.\eea If $\tan\beta\sim 1$, the above LL and RR
bounds can be satisfied even when $m_{\tilde{l}}\sim 300$ GeV,
$|\Delta^{l,e}_1-\Delta^{l,e}_2|\sim 1$, and
$\big(V_{L,R}^e\big)_{12}$ have generic values  satisfying the above
relation. However, for large $\tan\beta$, the $\mu\to e\gamma$ bound
requires a small $\left|\big(V_L^e\big)_{12}\right|$
unless $m_{\tilde{l}}\gg 300$ GeV or $|\Delta^{l,e}_1-\Delta^{l,e}_2|\ll 1$.
For the case with $m_{\tilde{l}}\sim 300$ GeV and
$|\Delta^{l,e}_1-\Delta^{l,e}_2|\sim 1$, which is actually the case
of interest for us, the $\mu\to e\gamma$ bound can be satisfied with
the following small mixing angle pattern as long as
$\tan\beta\lesssim 30$:
\bea\left|\big(V_L^e\big)_{12}\right|\,\sim\, \theta_C^n,\quad
\left|\big(V_R^e\big)_{12}\big)\right|\,\sim\,
\theta_C^{3-n}, \quad (n=1,2),\eea where $\theta_C\sim 0.2$ is the
Cabbibo angle. In this case, the large neutrino mixing angles in the
PMNS matrix $V_{\rm PMNS}$ should originate from the unitary matrix
$V_L^\nu$ diagonalizing the neutrino mass matrix as
(\ref{diagonalization}), and this might provide a nontrivial
condition on the mechanism to generate the neutrino masses. It is
interesting to note that this lepton mixing angle pattern allows a
sizable SUSY contribution to the muon anomalous magnetic moment,
\cite{Bennett:2006fi, Stockinger:2006zn, Hagiwara:2006jt}
which is given by \cite{Moroi:1995yh}\bea \frac{a_\mu^{\rm SUSY}}{1\times
10^{-9}}\simeq \left(\frac{\tan\beta}{6}\right) \left(\frac{300 {\rm
GeV}}{m_{\tilde{l}}}\right)^2\left(\frac{\mu}{m_{\tilde{l}}}\right)\eea
for a Wino mass $m_{\tilde{W}}\sim m_{\tilde{l}}$.

To summarize the flavor and CP constraints on moduli-mediated SUSY
breaking in flux compactification, we find that most of constraints
other than  those from $\epsilon_K$ and $\mu\to e\gamma$ are well
satisfied even for generic form of higher order K\"ahler potential
and sparticle spectra in the sub-TeV range, if the size of higher
order K\"ahler potential in the messenger modulus expansion is of
${\cal O}(g_{GUT}^2/8\pi^2)$. The constraints from $\epsilon_K$ and
$\mu\to e\gamma$ can be satisfied  also again for generic form of
higher order K\"ahler potential and sparticle spectra in the sub-TeV
range, if one makes a plausible assumption on flavor mixing angles
motivated by the observed hierarchical structure of quark and
charged lepton masses, for instance $|(V_L^d)_{12}(V_R^d)_{12}|\sim
m_d/m_s$ and $|(V^e_L)_{12}(V^e_R)_{12}|\sim m_e/m_\mu$ with
$|(V^e_L)_{12}|\lesssim 1/\tan\beta$.

\section{conclusion}

Flux compactification can provide a SUSY breaking scheme in which
soft terms preserve flavor and CP at leading order in the
perturbative expansion controlled by the vacuum expectation value of
the messenger modulus. In this paper, we have discussed some
features of flux compactification leading to such SUSY breaking
scheme, and examined the flavor and CP constraints on the higher
order K\"ahler potential. It is found that all phenomenological
constraints can be satisfied even for generic form of higher order
K\"ahler potential and sparticle spectra in the sub-TeV range, under
plausible  assumptions on the size of higher order correction and
flavor mixing angles. This implies that various SUSY breaking
schemes involving such modulus mediation, e.g. mirage mediation and
modulus-dominated mediation realized in flux compactification, can
be free from the SUSY flavor and CP problems, while giving gaugino
and sfermion masses in the sub-TeV range which can be probed by the
LHC.

\vskip 1cm
{\bf Acknowledgement}
\vskip 0.5cm
We thank T. Kobayashi for discussions which have motivated this
study. This work was supported by the Korea Research Foundation
Grant funded by the Korean Government (MOEHRD, Basic Research
Promotion Fund) (KRF-2005-210-C000006), the Center for High Energy
Physics of Kyungpook National University, and the BK21 program of
Ministry of Education. K.O. is supported by the Grant-in-aid for
Scientific Research No. 18071007 and No. 19740144 from the Ministry
of Education, Culture, Sports, Science and Technology of Japan.
\vskip 1cm

{\bf Appendix A: Matter modular weights}

\vskip 0.5cm
In some class of compactification,  the modular weights
can be determined by a simple scaling argument combined with the
non-linear PQ symmetry of the axion component
\cite{burgess,conlon1,Choi:2006za}. In our notation, the modular
weight $n_I$ is defined by the matter kinetic function as \bea Y_I
\propto(T+T^*)^{n_I}\eea at leading order in the messenger modulus
expansion. Note that \bea Y_I=e^{-K_0/3}Z_I,\eea where $K_0$ is the
moduli K\"ahler potential and $Z_I$ is the matter K\"ahler metric,
i.e \bea K=K_0(T+T^*)+Z_I(T+T^*)Q^{I*}Q^I.\eea At leading order,
$e^{-K_0}$ and $Z_I$  have a simple power-dependence on ${\rm
Re}(T)$: \bea \label{kahlerscaling} e^{-K_0}\propto
(T+T^*)^{n_0},\quad Z_I\propto (T+T^*)^{k_I},\eea and
then\footnote{Often  $-k_I$ is also called the matter modular
weight.} \bea n_I=\frac{1}{3}n_0+k_I.\eea

Typically, the messenger modulus behaves (in the string unit with
$\alpha^\prime=1$) as \bea {\rm Re}(T)\propto R^l/g_{st}^n,\eea
 where $g_{st}$
is the string coupling, $R$ is the compactification radius, and $l$
and $n$ are (model-dependent) non-negative integers. The string
coupling and compactification radius define another modulus $\propto
R^{l^\prime}/g_{st}^{n^\prime}$ which might be fixed by flux. Here,
we consider two simple cases: the case (A) with $n^\prime=0$, in
which $R$ is stabilized by flux, while $g_{st}$ remains unfixed,
and another case (B) with $l^\prime=0$, in which $g_{st}$ is
stabilized by flux, while $R$ remains unfixed. In case (A),  the
messenger modulus expansion can be identified as a string coupling
expansion with $g_{st}^n\propto 1/{\rm Re}(T)$. On the other hand,
in case (B), the messenger modulus expansion can be identified as
a radius expansion with $1/ R^l\propto 1/{\rm Re}(T)$.

\subsubsection{Case (A)}

Let us first examine the case that the messenger modulus expansion
corresponds to a string coupling expansion with \bea{\rm
Re}(T)\propto 1/g_{st}^n,\eea where $n$ is a positive integer. For
this case, we assume that the kinetic terms of 4D gauge and matter
fields and also the trilinear Yukawa couplings are generated at the
same (leading) order in $g_{st}$,and thus the $g_{st}$-dependence of
the 4D action is schematically given by \bea {\cal
L}=\frac{1}{g_{st}^N}\left[-\frac{1}{4}F^a_{\mu\nu}F^{a\mu\nu}+\partial_\mu
\phi^{I*}\partial^\mu\phi^I+i\bar{\psi}^I\sigma^\mu\partial_\mu\psi^I
+\left(\lambda_{IJK}\phi^I\psi^J\psi^K + {\rm h.c.}\right)\right],
\eea where $(\phi^I,\psi^I)$ denote the chiral matter multiplets,
$N$ is a positive integer, and $\lambda_{IJK}$ are independent of
$g_{st}$. We then have \bea g_{GUT}\propto g_{st}^{N/2}, \quad
y_{IJK}\propto g_{st}^{N/2}, \eea where  $g_{GUT}$ and $y_{IJK}$ are
the 4D gauge coupling and the canonically normalized Yukawa
couplings, respectively.

In $N=1$ superspace, this 4D action can be written as \bea \int
d^4\theta \,Y_IQ^IQ^{I*}+\left[ \int d^2\theta
\left(\frac{1}{4}f_aW^aW^a +\frac{1}{6}\lambda_{IJK}Q^I Q^J Q^K
\right) + {\rm h.c.} \right],\eea where
$Q^I=\phi^I+\theta\psi^I+\theta^2F^I$. The non-linear PQ symmetry
$U(1)_T$ of the axion component ${\rm Im}(T)$ implies that the
holomorphic Yukawa couplings $\lambda_{IJK}$ are independent of $T$,
while the gauge kinetic functions $f_a$ are either linear in $T$ or
independent of $T$. Combining those constraints from $U(1)_T$ with
\bea \frac{1}{g_{GUT}^2}={\rm Re}(f_a), \quad
y_{IJK}=\frac{\lambda_{IJK}}{\sqrt{Y_IY_JY_K}},\eea one easily finds
$N=n$, and \bea f_a\propto T, \quad Y_I \propto (T+T^*)^{n_I}\eea
with \bea n_I+n_J+n_K=1 \eea at leading order in the messenger
modulus expansion. On the other hand, the universal
$g_{st}$-dependence of matter kinetic terms and Yukawa couplings
suggests that the $T$-dependence of $Y_I$ is universal also, so \bea
n_I=1/3.\eea To summarize, if the messenger modulus expansion
corresponds to a string coupling expansion, and the gauge and matter
kinetic terms and the trilinear Yukawa couplings are generated at
the same (leading) order in this expansion, the matter modular
weights have a universal value $1/3$. One such example  is the  case
that the messenger modulus corresponds to the heterotic dilaton, for
which $n_0=1$ and $k_I=0$ in (\ref{kahlerscaling}), and thus
$n_I=1/3$.

Quite often, string compactification involves an anomalous $U(1)_A$
gauge symmetry \cite{anomalousu1} under which $T$ transforms as \bea
U(1)_A: \,T\,\rightarrow\, T-\frac{i}{2}\alpha(x)\delta_{GS},\eea
where $\alpha(x)$ is the $U(1)_A$ transformation function and
$\delta_{GS}$ is the Green-Schwarz coefficient of ${\cal
O}(1/8\pi^2)$. In the presence of such anomalous $U(1)_A$,
 the messenger modulus should be redefined as it mixes with
the $U(1)_A$ vector superfield $V$. This results in a shift of
modular weight  after the massive $U(1)_A$ vector multiplet is
integrated out, as will be discussed below.

Models with anomalous $U(1)_A$ give a modulus-dependent
Fayet-Iliopoulos (FI) D-term \bea
\xi_{FI}=\frac{1}{2}\delta_{GS}\frac{\partial K_0}{\partial T}.\eea
Then, to satisfy the D-flat condition, one needs a $U(1)_A$-charged
MSSM singlet $X$ which has a large vacuum value $\langle
X\rangle={\cal O}(\xi_{FI})$ to cancel this FI term.

Let us consider the 4D action including such field $X$: \bea {\cal
L}&=&\int d^4\theta\, \left[\,\Omega_0(T+T^*-\delta_{GS}V)+
Y_X(T+T^*-\delta_{GS}V)X^*e^{-2V}X\right.\nonumber
\\
&&\left.+\,Y_I(T+T^*-\delta_{GS}V)Q^{I*}e^{2q_I V}Q^{I}\right],\eea
where $\Omega_0\equiv -3e^{-K_0/3}$ and  the $U(1)_A$ charge of $X$
is normalized as $q_X=-1$. For $\delta_{GS}={\cal O}(1/8\pi^2)$, one
can show \cite{choi-u1a} that the mass eigenstate vector superfield
$\tilde{V}$ is given  by  \bea \tilde{V}\simeq V-\ln|X|\eea which
has a superheavy mass $M_{\tilde{V}}^2\sim \delta_{GS}M_{Pl}^2$. It
is straightforward to integrate out $\tilde{V}$ to obtain the
effective action of the light modulus $T$ and the visible matter
fields $Q^i$: \bea {\cal L}_{\rm eff}= \int d^4\theta\,\left[\,
\Omega_0(T+T^*)+Y_I^{\rm eff}(T+T^*)Q^{I*}Q^I+...\right],\eea where
the ellipsis stands for the corrections suppressed by $\delta_{GS}$,
and the effective matter kinetic function is given by (after an
appropriate redefinition of $Q^I$) \cite{dine,choi-u1a} \bea
\label{shift} Y_I^{\rm
eff}=\left(\frac{Y_X}{\partial_T\Omega_0}\right)^{q_I}Y_I.\eea

After $\tilde{V}$ is integrated out, the effective modular weight is
defined as \bea Y_I^{\rm eff}\propto (T+T^*)^{n_I^{\rm eff}}\eea at
leading order in the messenger modulus expansion. In case when $T$
corresponds to the heterotic dilaton, we have
$\Omega_0,Y_{X},Y_I\propto (T+T^*)^{1/3}$. The  resulting effective
modular weight is give by \bea n_I^{\rm eff}=\frac{1}{3}+q_I,\eea
which would be flavor universal if the $U(1)_A$ charges are flavor
universal.

\subsubsection{Case (B)}

Let us consider another case that the messenger modulus expansion
corresponds to a radius expansion with \bea {\rm Re}(T)\propto R^l
\quad (l>0).\eea In this case, we can have  more variety of
possibilities.

Let us suppose that the gauge field $A^a_\mu$ propagates over
$l_G$-dimensional internal space ($l_G>0$), the matter field $Q^I$
propagates over  $l_I$-dimensional internal space, and the Yukawa
coupling $y_{IJK}$ originates from a wavefunction integral over
$l_{IJK}$-dimensional internal space.  Then, schematically, the 4D
action takes the form: \bea{\cal
L}=-\frac{1}{4}R^{l_G}F^a_{\mu\nu}F^{a\mu\nu}+R^{l_I}\left(\partial_\mu
\phi^{I*}\partial^\mu\phi^I+i\bar{\psi}^I\sigma^\mu\partial_\mu\psi^I\right)
+\left(R^{l_{IJK}}\lambda_{IJK}\phi^I\psi^J\psi^K + {\rm
h.c.}\right),\eea where \bea 0\leq l_I\leq l_G,\quad 0\leq
l_{IJK}\leq \mbox{min}(l_I,l_J,l_K).\eea The resulting gauge and
canonically normalized Yukawa couplings behave as \bea
\frac{1}{g_{GUT}^2}&=&{\rm Re}(f_a)\,\propto\, R^{l_G},
\nonumber \\
y_{IJK}&=& \frac{\lambda_{IJK}}{\sqrt{Y_IY_JY_K}}\,\propto\,
R^{l_{IJK}-\frac{l_I+l_J+l_K}{2}}.\eea

Again, with the non-linear PQ symmetry $U(1)_T$ which requires $f_a$
is either linear in $T$ or independent of $T$, and $\lambda_{IJK}$
are independent of $T$,  these relations imply $l_G=l$, and \bea
Y_I\propto (T+T^*)^{n_I},\eea where $n_I$ are constrained as \bea
n_I+n_J+n_K=\frac{l_I+l_J+l_K-2l_{IJK}}{l_G}.\eea For the MSSM
matter fields, it is quite plausible that $l_I$ and $l_{IJK}$ are
universal.  Then, the resulting modular weights are universal and
given by \bea
n_I=\frac{l_I}{l_G}-\frac{2}{3}\frac{l_{IJK}}{l_G}.\eea One
interesting point is that the modular weights have a universal value
1/3 as in the case (A) if all gauge and matter fields propagate over
the same internal space and also the Yukawa couplings are given by
wavefunction integrals over the same internal space, i.e.
$l_G=l_I=l_{IJK}$. In models with an anomalous $U(1)_A$, this
modular weight is shifted by the $U(1)_A$ charge as determined by
(\ref{shift}).

\end{document}